\documentclass[namedreferences]{SolarPhysics}
\usepackage[optionalrh]{spr-sola-addons} % For Solar Physics
\usepackage{graphicx}                    % For eps figures, newer & more powerfull
\usepackage{color}                       % For color text: \color command
\usepackage{url}                         % For breaking URLs easily trough lines
\begin{document}

\begin{article}

\begin{opening}

\title{Time-Distance Solar Far-Side Imaging Using Three-Skip Acoustic Signals}

%%%%%%%%%%%%%%%%%%%%%%%%%%%%%%%%%%%%%%%%%%%%%%%%%%%
%% Authors Names
%
\author{S.~\surname{Ilonidis}$^{1}$\sep
        J.~\surname{Zhao}$^{1}$\sep
        T.~\surname{Hartlep}$^{2,3}$
       }

%%%%%%%%%%%%%%%%%%%%%%%%%%%%%%%%%%%%%%%%%%%%%%%%%%%
%% Runningheads
%
\runningauthor{S. Ilonidis \emph{et al.}}
\runningtitle{Time-Distance Far-Side Imaging}

%%%%%%%%%%%%%%%%%%%%%%%%%%%%%%%%%%%%%%%%%%%%%%%%%%%
%% Affilations
%

   \institute{$^{1}$ W. W. Hansen Experimental Physics Laboratory, Stanford University,
Stanford, CA 94305-4085, USA \\ email: \url{ilonidis@sun.stanford.edu} \\
$^{2}$ NASA Ames Research Center, MS 230-2, Moffett Field, CA 94035-1000, USA \\
%email: \url{thartlep@sun.stanford.edu} \\
%$^{3}$ Center for Turbulence Research, Stanford University,
%Stanford, CA 94305-3035
$^{3}$ present address: W. W. Hansen Experimental Physics
Laboratory, Stanford University, Stanford, CA 94305-4085, USA
              }

%  \institute{S. Ilonidis \sep J. Zhao \\
%W. W. Hansen Experimental Physics Laboratory, Stanford University,
%Stanford, CA 94305-4085.
%                     email: \url{ilonidis@stanford.edu} \\
%                     J. Zhao \\
%                     email: \url{junwei@sun.stanford.edu} \\
%                     T. Hartlep \\ NASA Ames Research
%Center, MS 230-2, Moffett Field, CA 94035-1000.
%                     email: \url{thomas.hartlep@nasa.gov} \\
%                     T. Hartlep \\  Center for
%Turbulence Research, Stanford University, Stanford, CA 94305-3035.
%             }

%%%%%%%%%%%%%%%%%%%%%%%%%%%%%%%%%%%%%%%%%%%%%%%%%%%
%%% Abstract
\begin{abstract}
The purpose of this work is to image solar far-side active regions
using acoustic signals with three skips and improve the quality of
existing images. The mapping of far-side active regions was first
made possible using the helioseismic holography technique by use of
four-skip acoustic signals. The quality of far-side images was later
improved with the combination of four- and five-skip signals using
the time-distance helioseismology technique. In this work, we
explore the possibility of making three-skip far-side images of
active regions, and improving the image quality by combining the
three-skip images with the images obtained from existing techniques.
A new method of combining images is proposed that increases the
signal-to-noise ratio and reduces the appearance of spurious
features.
\end{abstract}

%%%%%%%%%%%%%%%%%%%%%%%%%%%%%%%%%%%%%%%%%%%%%%%%%%%
%% Keywords
%
\end{opening}
%-------------------------------------------------

%%%%%%%%%%%%%%%%%%%%%%%%%%%%%%%%%%%%%%%%%%%%%%%%%%%
%% Sections
%
 \section{Introduction}%\label{s:?}
\par The detection of large solar active regions on the far side of
the Sun can greatly improve space-weather forecasting and facilitate
the study of evolution of those active regions. Active regions that
appear on the east solar limb can affect space weather, causing
problems in spacecraft, electrical power grids, and
telecommunications. Monitoring the solar far-side activity allows
for the anticipation of large active regions by more than a week
before they rotate into our view from the Sun's east limb.
\par The first attempt to map the central
region of the solar far side was made by Lindsey and Braun (2000a)
by use of the helioseismic holography technique (Lindsey and Braun,
2000b). Low- and medium-$\ell$ acoustic waves exhibit an apparent
travel time deficit in solar active regions which is detectable by
phase-sensitive holography (Braun and Lindsey, 2000). This technique
initially included only double-skip acoustic signals but was later
extended (Braun and Lindsey, 2001) to map the whole far side using
single- and triple-skip signals.
\par The time-distance helioseismology technique also has the capability to map the central
area of the solar far side (Duvall, Kosovichev, and Scherrer, 2000;
Duvall and Kosovichev, 2001). The mapping area was later extended
onto the whole far side using four- and five-skip acoustic signals
(Zhao, 2007). The two schemes are used separately to make two whole
far-side maps. The combination of the two maps gives far-side images
with better signal-to-noise ratio. This is an independent solar
far-side imaging tool in addition to the holography technique which
allows for the possibility to cross-check the active regions seen in
both time-distance and helioseismic holography techniques and hence
improves the accuracy of monitoring the solar far-side active
regions.
\par However, no studies have been carried
out yet to explore the possibility of making far-side images using
acoustic signals with three skips. In this paper we apply the
time-distance technique to show that it is possible to image the
solar far side utilizing three-skip acoustic signals, and discuss
advantages and disadvantages of this method. We also suggest a
method to improve the quality of existing far-side images that
includes the results of our work.

\section{Data and Technique}%\label{s:?}

\par The medium-$\ell$ program $\!$ of the \emph{Solar and Heliospheric
Observatory} / Michelson Doppler Imager (SOHO/MDI) (Scherrer
\emph{et al.}, 1995) is used in this work. The oscillation modes
that are observed range from $\ell=0$ to $\thickapprox 300$ and the
data are acquired with one-minute cadence and a spatial sampling of
$10''$ (0.6 heliographic degrees per pixel). The $\ell-\nu$ power
spectrum diagram of this dataset is shown in Figure 1a. The
continuity of the observations and the particular range of modes
that we analyze make the medium-$\ell$ data ideal for solar far-side
imaging.

\par The time-distance technique is able to detect acoustic signals
that travel to the far side and return to the front side after
experiencing four and five skips (Zhao 2007). It is useful to
examine whether acoustic signals after three skips are also
detectable by this technique. In Figure 1b the time-distance diagram
is computed using all oscillation modes from a 2048-minute MDI
medium-$\ell$ dataset with a spatial size of $120^\circ \times
120^\circ$ after the region is tracked with the Carrington rotation
rate and is remapped using Postel's projection with the solar disk
center as the remapping center. The diagram shows acoustic skips of
up to seven times on the front side as well as signals traveling to
the far side and returning to the front side after three and four
skips. From all of these signals we keep only the ones that are
needed in our analysis and filter out all other acoustic modes. The
acoustic signals that are needed should have long single-skip travel
distance corresponding to low-$\ell$ modes. Such a filtering
improves the signal-to-noise ratio at the required travel distances
of the three-skip method. Figure 1c shows the time-distance diagram
computed using only the useful acoustic modes with a box of
$\ell$-range $6-15$ at 2.5 mHz and $12-30$ at 4.5 mHz. This
filtering is applied to the data after Postel's projection and thus
it works best where the geometry is least distorted. Another issue
is that the theoretical travel time is a few minutes off from the
time-distance group travel time after three skips, because as
pointed out by Kosovichev and Duvall (1996) and Duvall \emph{et al.}
(1997), the ray-approximated travel time is a few minutes off from
the time-distance group travel time.

\begin{figure}    %%%%%%%%%%%%%%%%%% FIGURE 2
                                % includes the two top panels
   \centerline{\hspace*{0.28\textwidth}
               \includegraphics[width=1.0\textwidth,clip=]{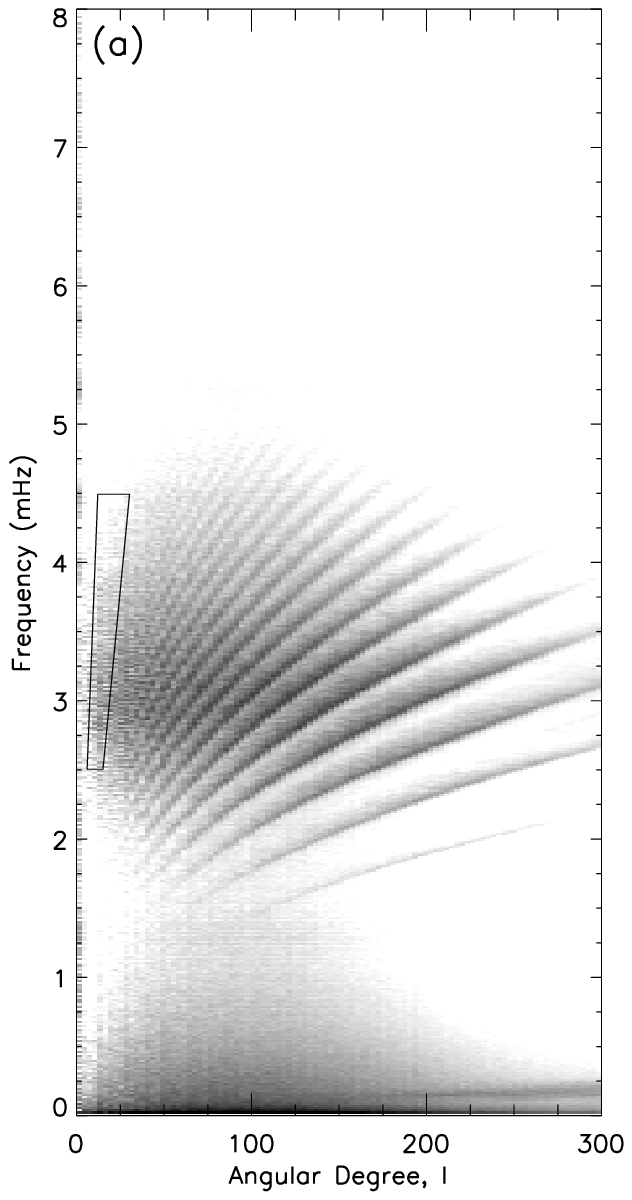}
               \hspace*{-0.94\textwidth}
               \includegraphics[width=1.0\textwidth,clip=]{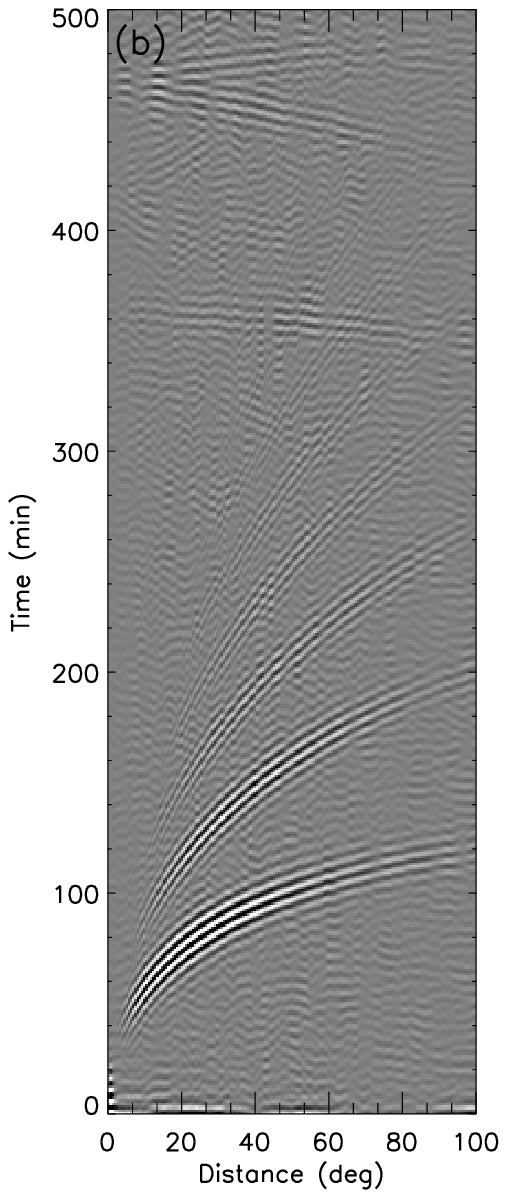}
               \hspace*{-0.71\textwidth}
               \includegraphics[width=1.0\textwidth,clip=]{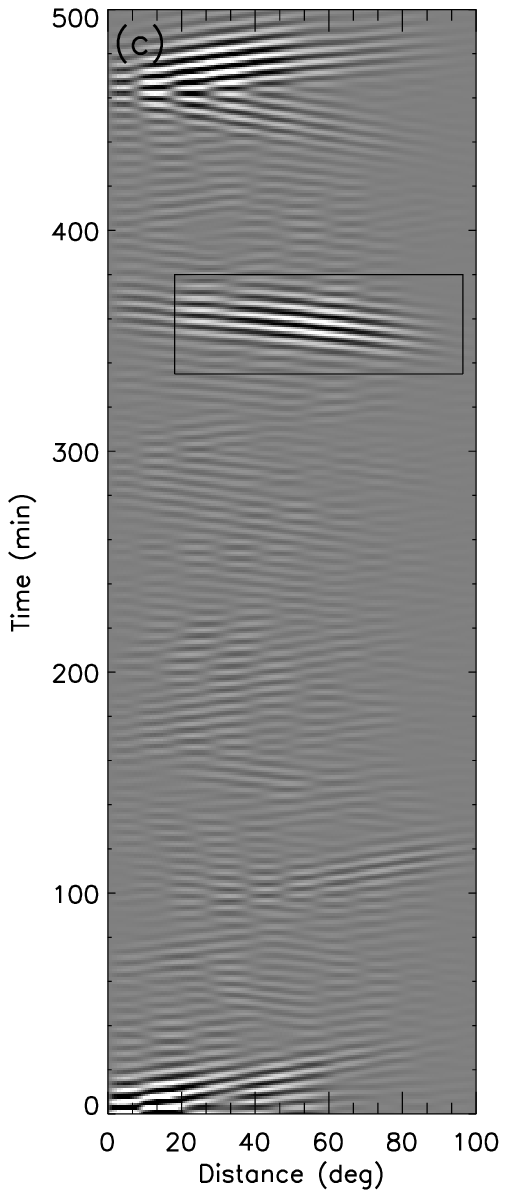}
              }
              \caption{(a) Power spectrum computed from a
              2048-minute MDI medium-$\ell$ dataset, (b) time-distance
              diagram computed using the whole power spectrum of the
              same dataset, and (c) time-distance diagram computed
              using only the oscillations that have frequency and
              $\ell$ included in the black quadrangle as indicated
              in (a). The black box delimits the acoustic travel
              distances and times used for far-side imaging.
                       }
   \label{F1}
   \end{figure}

\par Not all acoustic signals that come back
from the far side after three skips are used but only those with
time-distance annulus radii $80^\circ-113^\circ$ from the target
point for the single-skip and $160^\circ-226^\circ$ for the
double-skip (see Figure 2). This scheme is able to cover the whole
far side except for a circular region in the center of that side.
Following Zhao (2007) to save computation time, only the areas lower
than the latitude of 48$^\circ$, where nearly all active regions are
located, are included in the far-side imaging computations.

\begin{figure}
   \centerline{\includegraphics[width=0.8\textwidth,clip=]{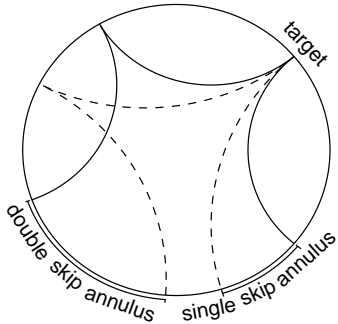}
              }
              \caption{Sketch for the three-skip  measurement scheme
              where one set of single-skip and one set of double-skip
              rays are located on the either side of the target
              point.
                       }
   \label{F2}
   \end{figure}

\par The first step of the computational procedure is to track a 2048-minute
long MDI medium-$\ell$ dataset with the Carrington rotation rate.
Then it is remapped to Postel's coordinates centered on the solar
disk with a spatial sampling of $0.6^\circ$ per pixel, covering a
span of $120^\circ$ along the Equator and the central meridian. This
dataset is filtered in the Fourier domain and only the oscillation
modes with travel distances in agreement with the distances listed
above are kept. Corresponding pixels in the annuli on both sides of
the target point are selected and the cross-covariances with both
positive and negative travel time lags are computed. Then both time
lags are combined and all cross-covariances obtained from different
distances after appropriate shifts in time are also combined. The
final cross-covariance is fitted with a Gabor wavelet function
(Kosovichev and Duvall, 1996) and an acoustic phase travel time is
obtained. The far-side image is a display of the measured mean
acoustic travel times after a Gaussian smoothing with a FWHM of
$2.0^\circ$.

\section{Results}
\subsection{Results from Numerical Simulation Data}
\par It is still not clear how accurate the far-side imaging
techniques are. Comparing the far-side images with the directly
observed near side after active regions have rotated into our view
from the far side, or before they rotate out of view onto the far
side, is the traditional way to evaluate the accuracy of far-side
images. However, active regions sometimes develop rapidly, on time
scales of days, thus the traditional way is not always sufficient.
Numerical models of solar oscillations can provide artificial data
with near-surface perturbations mimicking active regions on the far
side of the Sun. Applying helioseismic imaging techniques on the
simulated wavefield allows for a direct comparison of the far-side
images with the prescribed sound speed perturbations, and it thus
provides a more accurate test of the far-side imaging technique.

\par We present results on testing the new time-distance technique
with three-skip acoustic signals. The sensitivity of the technique
is tested by varying the size of the active region. The propagation
of solar acoustic waves is simulated numerically in a spherical
domain. The simulations take into account the effects of a spatially
varying sound speed but for now ignore flows or magnetic fields. The
quiet solar medium is represented with having a sound speed that is
only a function of radius given by the standard solar model S of
Christensen-Dalsgaard \emph{et al.} (1996) matched to a
chromospheric model from Vernazza, Avrett, and Loeser (1981). Active
regions are modeled as local variations in the sound speed from the
quiet Sun. The depth profile of these variations can be found in
Hartlep \emph{et al.} (2008). The data are prepared to be similar to
MDI medium-$\ell$ data and have been used to test the existing four-
and five-skip far-side imaging technique (Hartlep \emph{et al.},
2008). The radial velocity is computed at a location 300 km above
the photosphere and is stored with one-minute cadence and a spatial
resolution of $0.703^\circ$ pixel$^{-1}$. The data are remapped onto
Postel's coordinates with a spatial sampling of $0.6^\circ$
pixel$^{-1}$. A region of $120^\circ \times 120^\circ$ is used for
analysis and the first 500 minutes are discarded because they
represent transient behavior. The following 1024 minutes are used in
the analysis, a short but sufficient period for far-side analysis.
The rest of the procedure for the simulation data is the same as for
the observations. For more details about the simulation and for
discussion on how these simulations were used for testing far-side
imaging, see Hartlep \emph{et al.} (2008).

\begin{figure}    %%%%%%%%%%%%%%%%%% FIGURE 2
                                % includes the two top panels
   \centerline{\hspace*{0.015\textwidth}
               \includegraphics[width=0.515\textwidth,clip=]{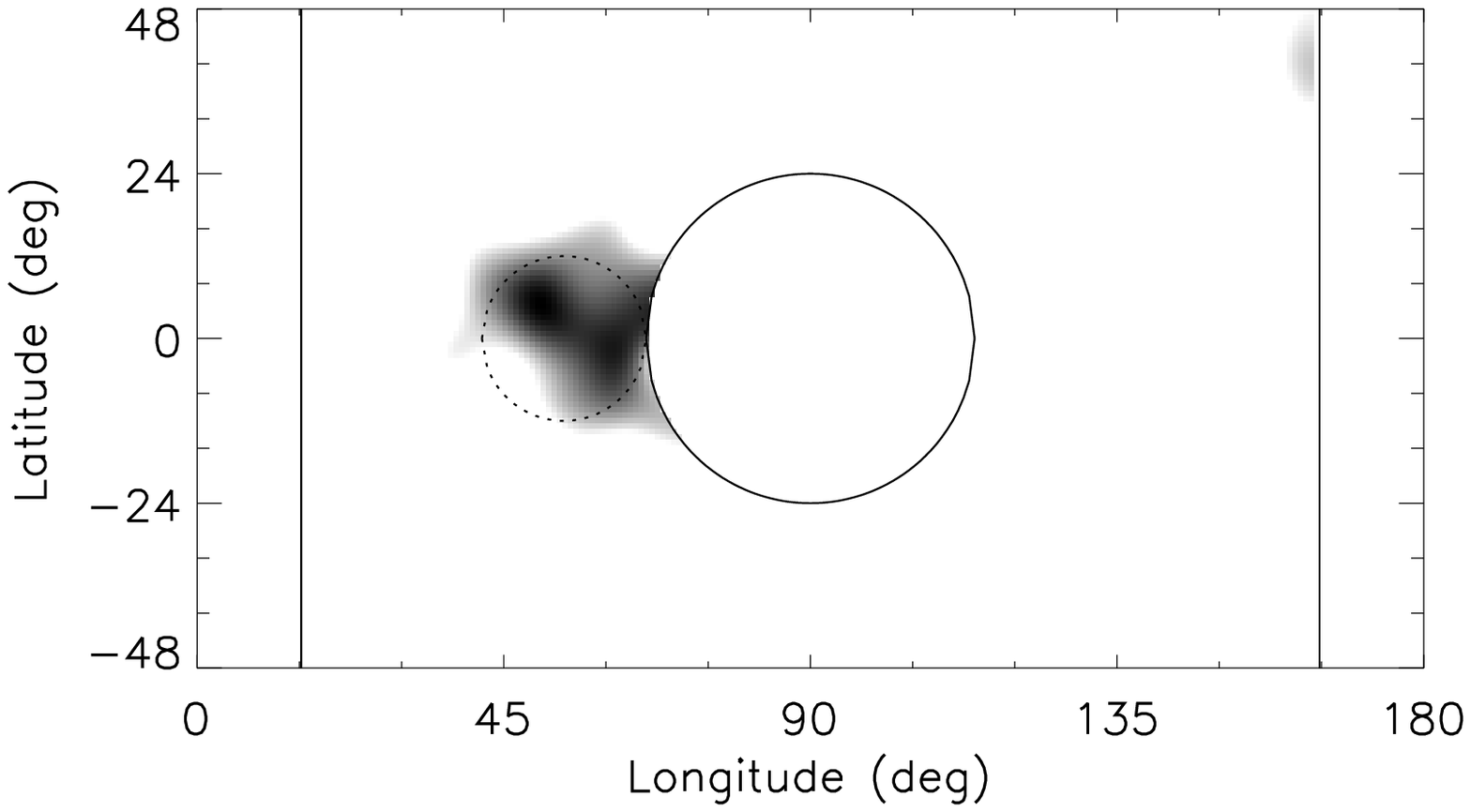}
               \hspace*{-0.03\textwidth}
               \includegraphics[width=0.515\textwidth,clip=]{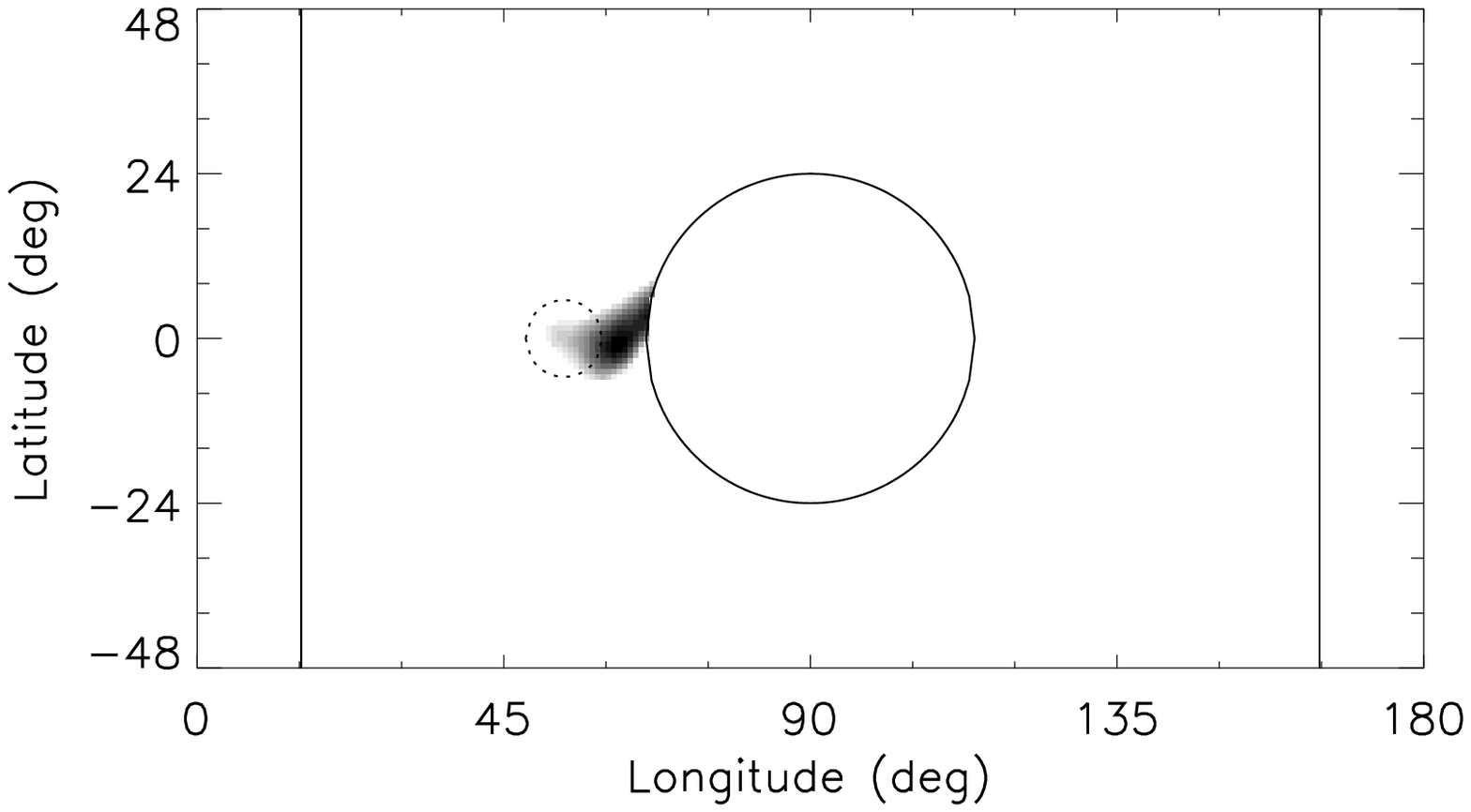}
              }
     \vspace{-0.35\textwidth}   % Shift close to the panel top
     \centerline{\Large \bf     % Includes the labels (here needs the color
                                %   package, see beginning of this file)
      \hspace{0.0 \textwidth}  \color{white}{(a)}
      \hspace{0.415\textwidth}  \color{white}{(b)}
         \hfill}
     \vspace{0.31\textwidth}    % Shift back to the panel bottom

\caption{Far-side images for simulations with an active region with
radius 180 Mm (left) and 90 Mm (right). The dotted circles indicate
the sizes and locations of active regions.
        }
   \label{F-2panels}
   \end{figure}

\par We have used simulated data for two active-region sizes, a large
one with a radius of 180 Mm and a smaller one with a radius of 90
Mm. A circular area around the center of the far side with a radius
of $24^\circ$ as well as a band along the solar limb $15^\circ$ in
width have been excluded from the far-side map because the
three-skip technique is not able to fully cover these areas. As seen
in the left image of Figure 3, the three-skip scheme can clearly
detect the large active region but with some level of uncertainty in
the location as well as the shape of the region. The detection of
the small active region by the three-skip scheme is shown in the
right image of Figure 3. It is quite remarkable that the far-side
map is completely clear of spurious features. However, as in the
previous case, the location and the shape are slightly different
from the original region.
\subsection{Results from MDI observations}
\par We have applied our three-skip imaging technique to the
numerical simulation data and found that this technique is able to
recover active regions larger than at least 90 Mm in size, but
unable to detect unambiguously features close to the far-side limbs
and center due to the low, or no, acoustic wave coverage in these
areas.

\begin{figure}    %%%%%%%%%%%%%%%%%% FIGURE 2
                                % includes the two top panels
   \centerline{\hspace*{0.015\textwidth}
               \includegraphics[width=0.515\textwidth,clip=]{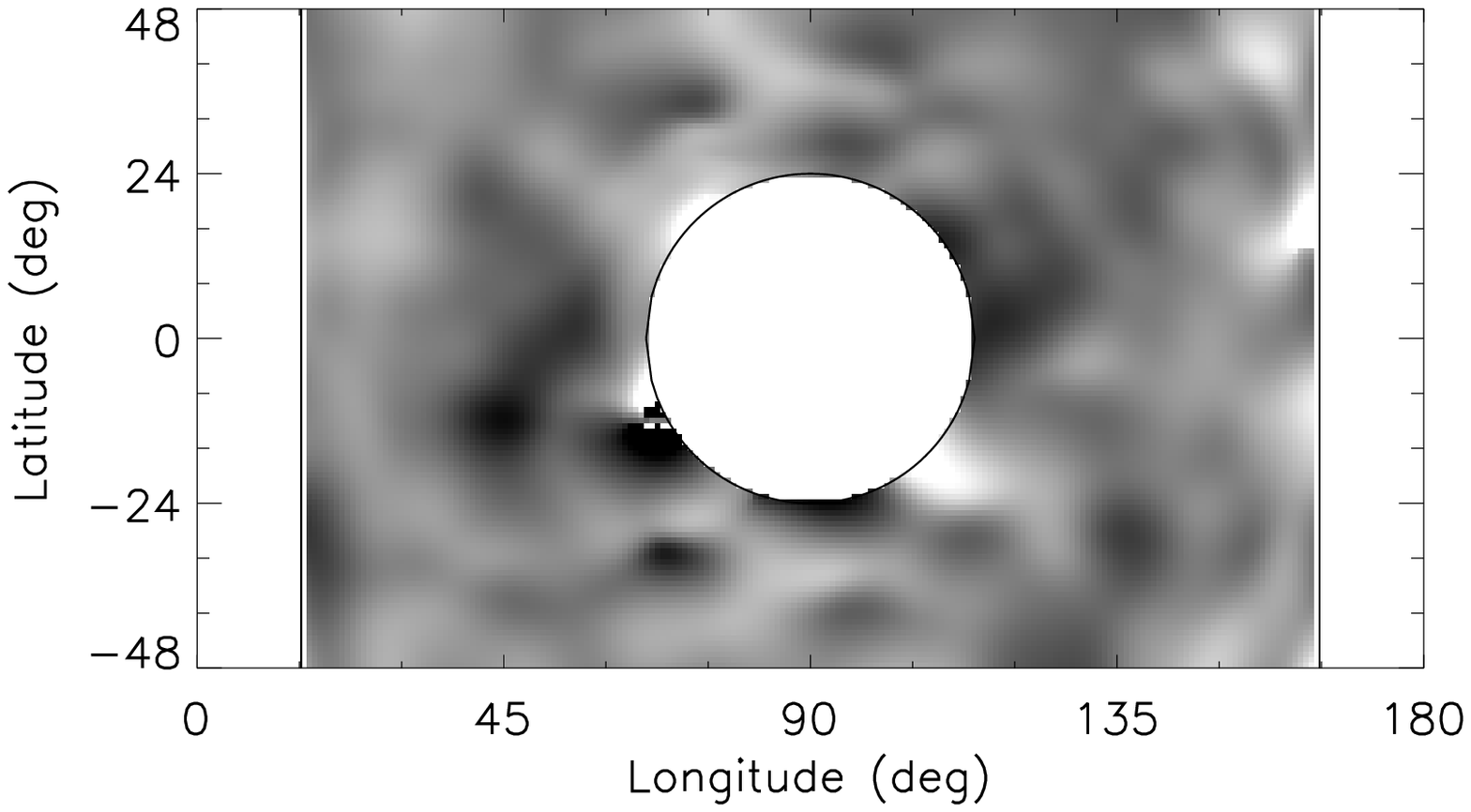}
               \hspace*{-0.03\textwidth}
               \includegraphics[width=0.515\textwidth,clip=]{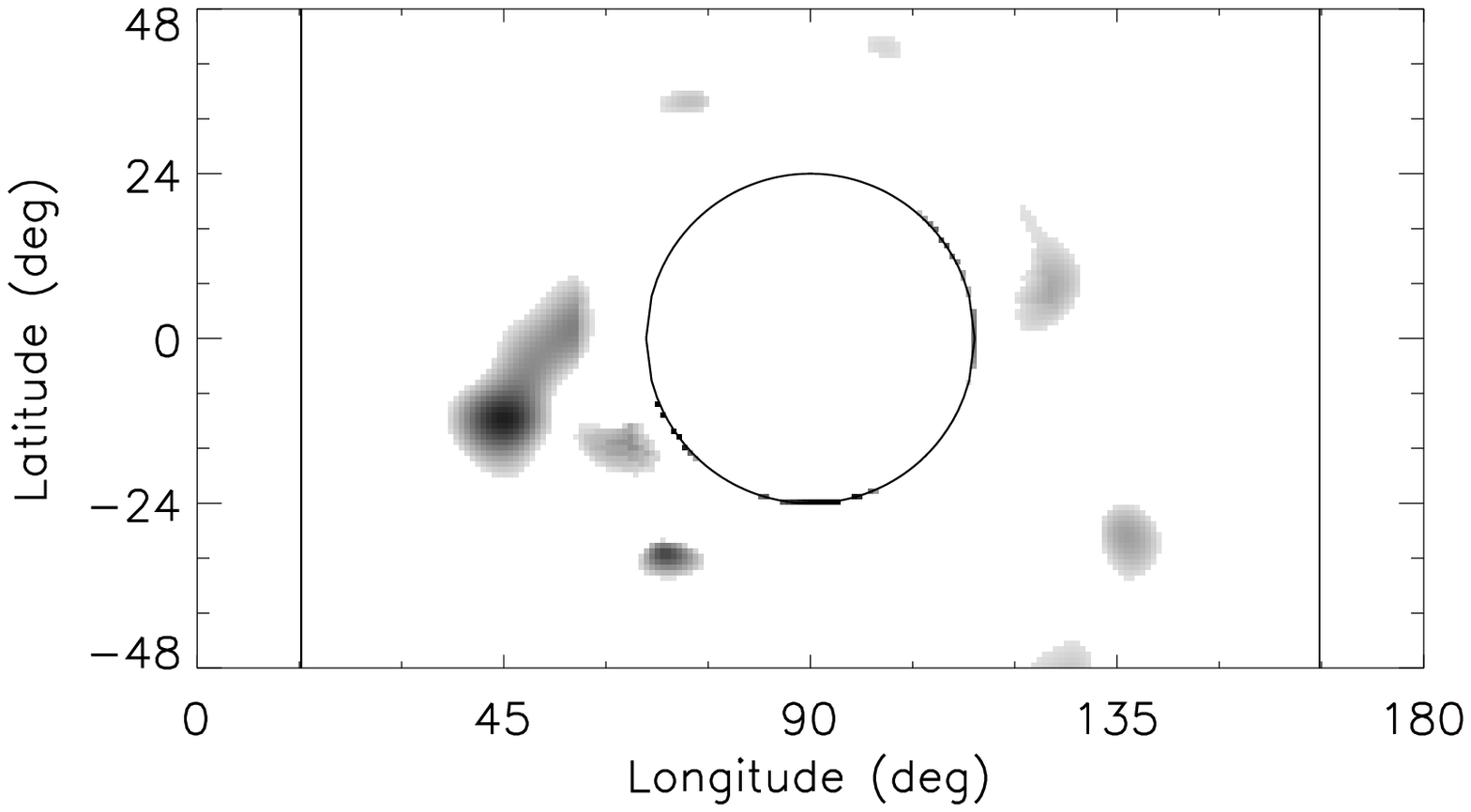}
              }
     \vspace{-0.35\textwidth}   % Shift close to the panel top
     \centerline{\Large \bf     % Includes the labels (here needs the color
                                %   package, see beginning of this file)
      \hspace{0.0 \textwidth}  \color{white}{(a)}
      \hspace{0.415\textwidth}  \color{white}{(b)}
         \hfill}
     \vspace{0.23\textwidth}    % Shift back to the panel bottom
   \centerline{\hspace*{0.015\textwidth}
               \includegraphics[width=0.515\textwidth,clip=]{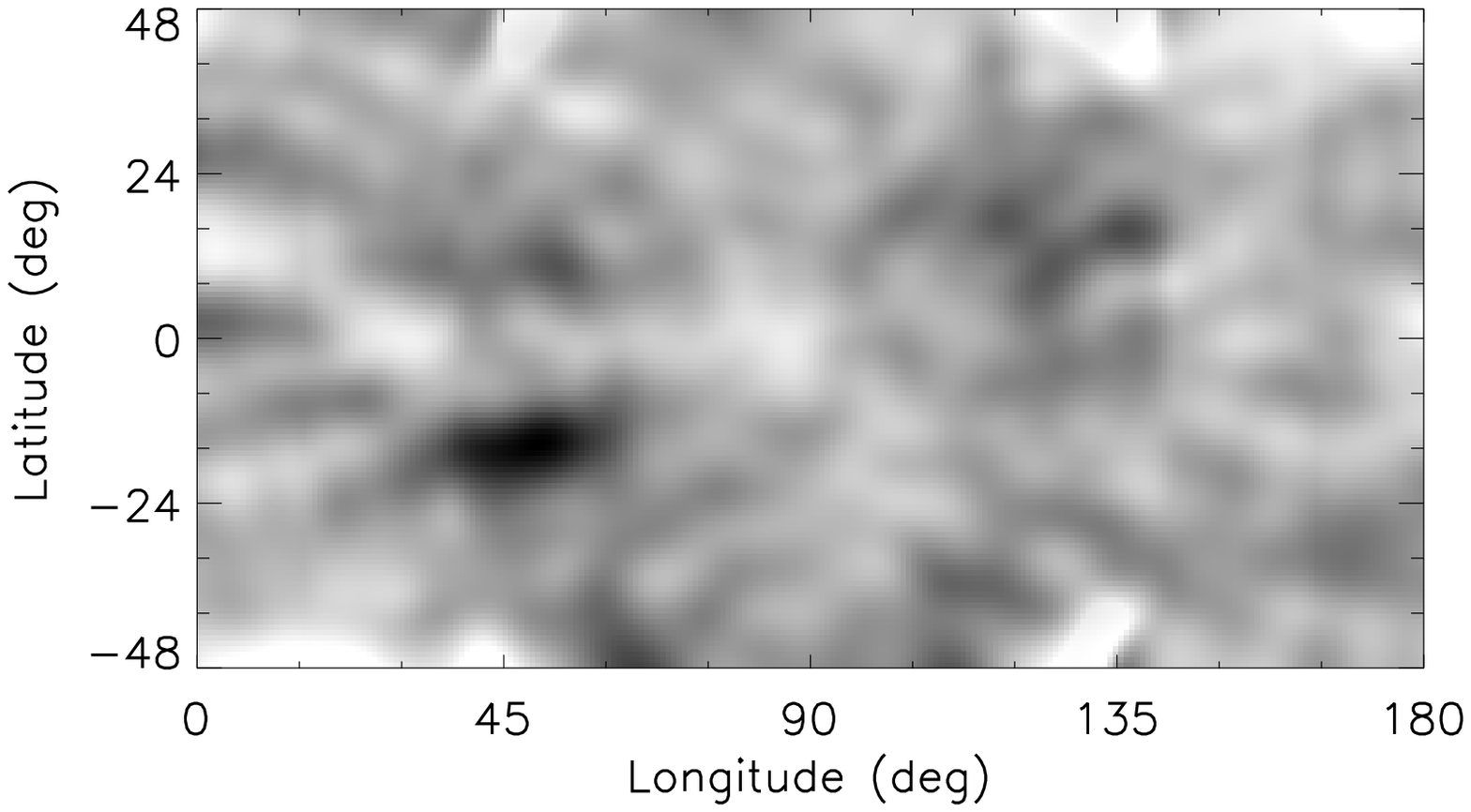}
               \hspace*{-0.03\textwidth}
               \includegraphics[width=0.515\textwidth,clip=]{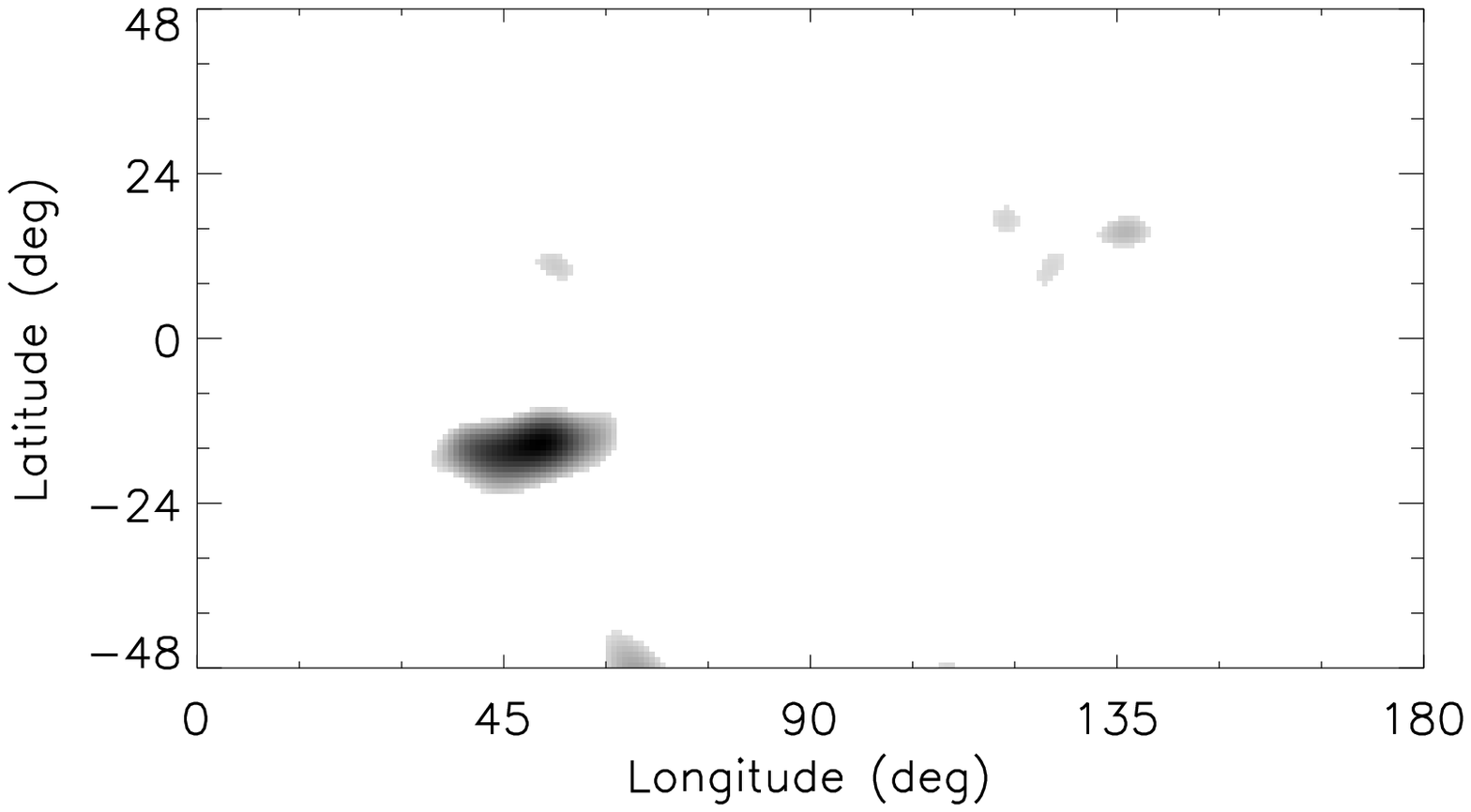}
              }
     \vspace{-0.35\textwidth}   % Shift close to the panel top
     \centerline{\Large \bf     % Includes the labels (here needs the color package)
      \hspace{0.0 \textwidth} \color{white}{(c)}
      \hspace{0.415\textwidth}  \color{white}{(d)}
         \hfill}
     \vspace{0.23\textwidth}    % Shift back to the panel bottom
     \centerline{\hspace*{0.015\textwidth}
               \includegraphics[width=0.515\textwidth,clip=]{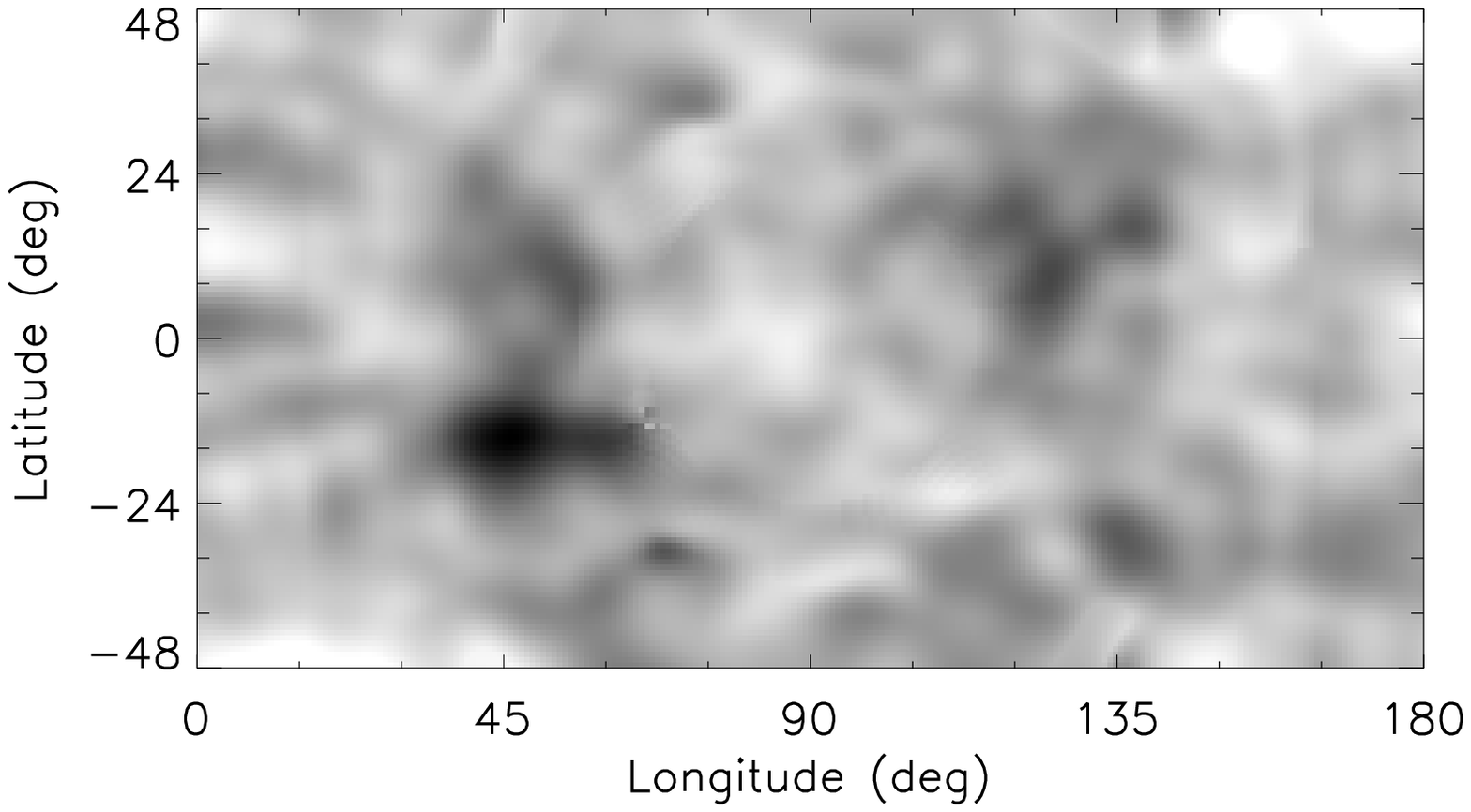}
               \hspace*{-0.03\textwidth}
               \includegraphics[width=0.515\textwidth,clip=]{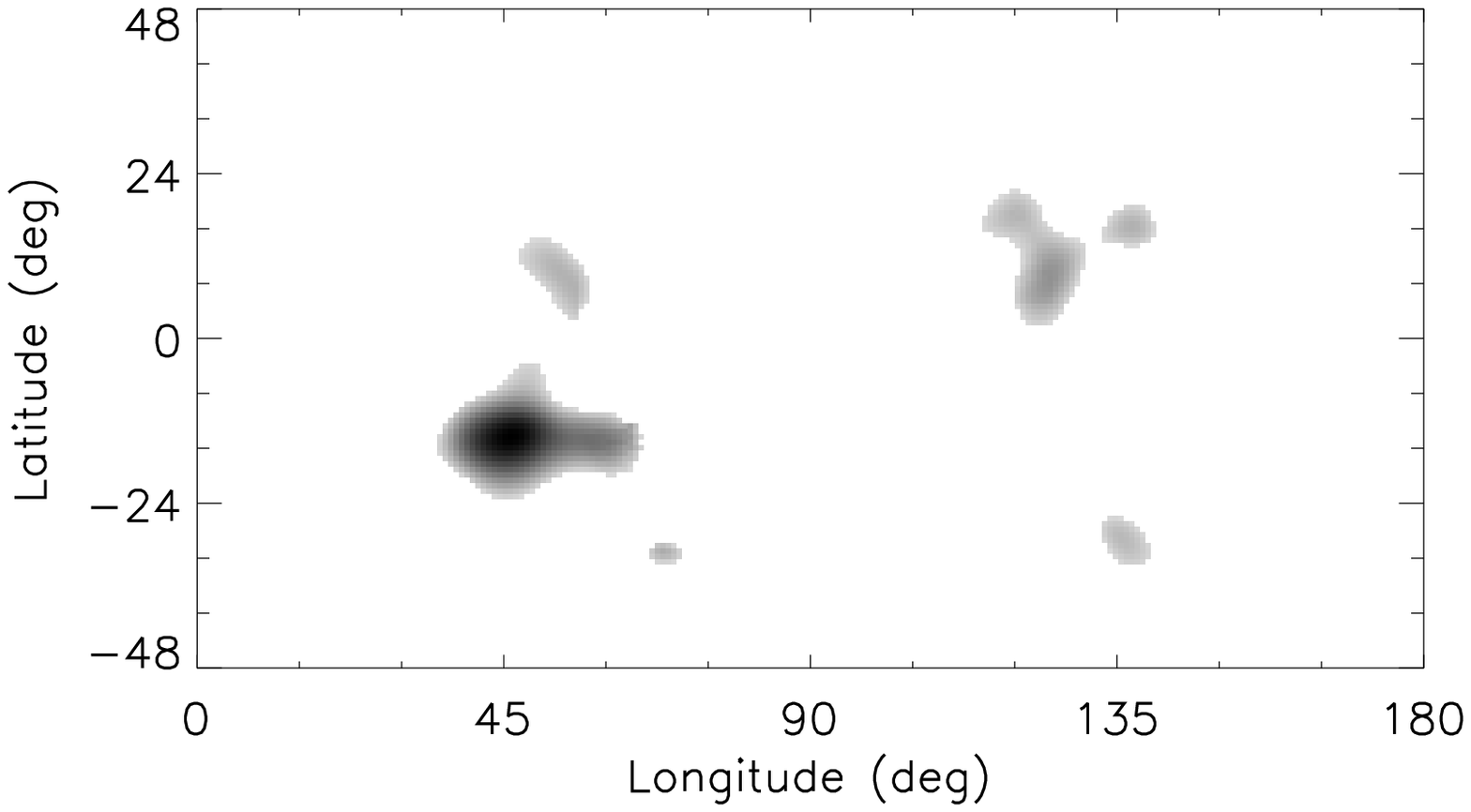}
              }
     \vspace{-0.35\textwidth}   % Shift close to the panel top
     \centerline{\Large \bf     % Includes the labels (here needs the color
                                %   package, see beginning of this file)
      \hspace{0.0 \textwidth}  \color{white}{(a)}
      \hspace{0.415\textwidth}  \color{white}{(b)}
         \hfill}
     \vspace{0.23\textwidth}    % Shift back to the panel bottom
   \centerline{\hspace*{0.015\textwidth}
               \includegraphics[width=0.515\textwidth,clip=]{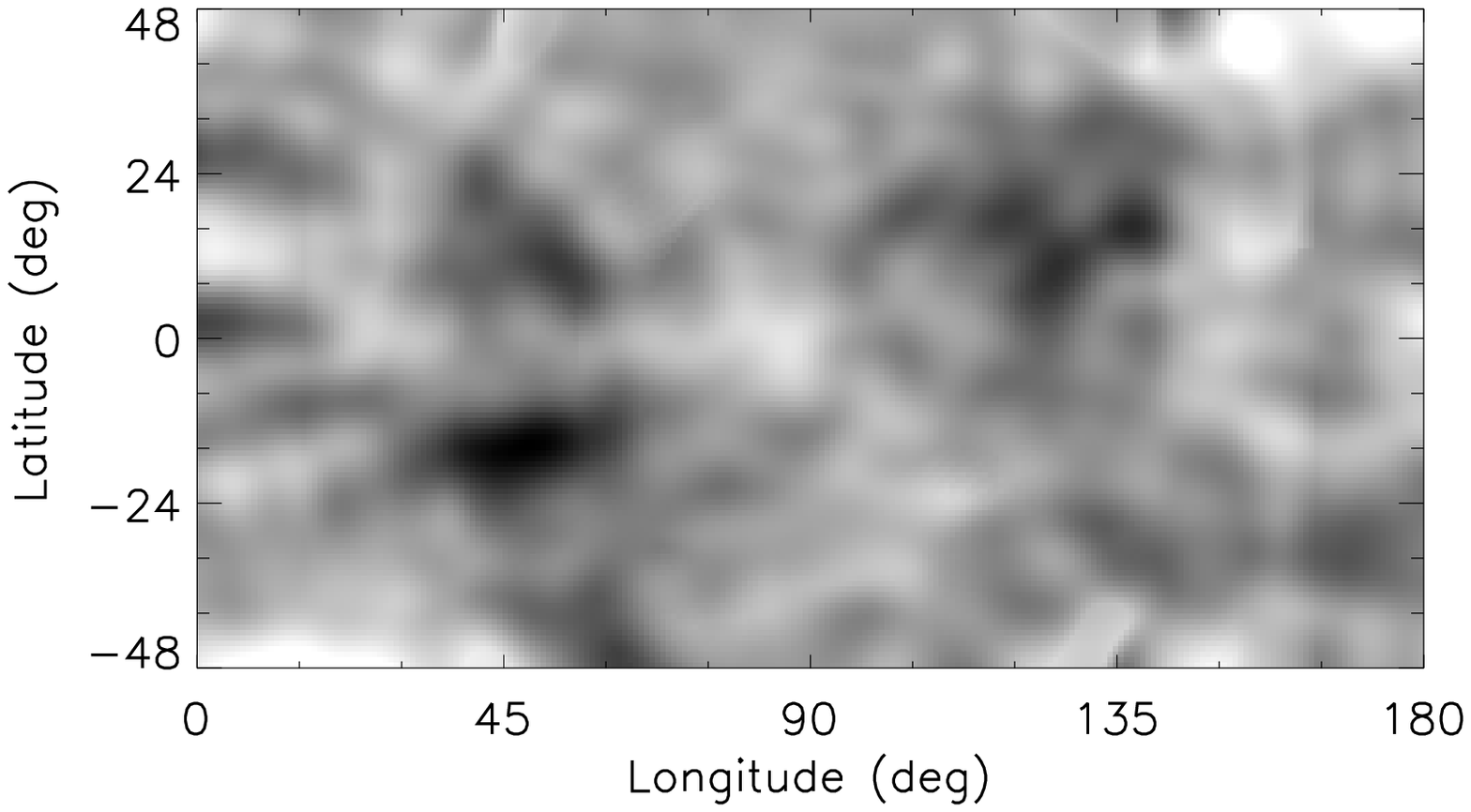}
               \hspace*{-0.03\textwidth}
               \includegraphics[width=0.515\textwidth,clip=]{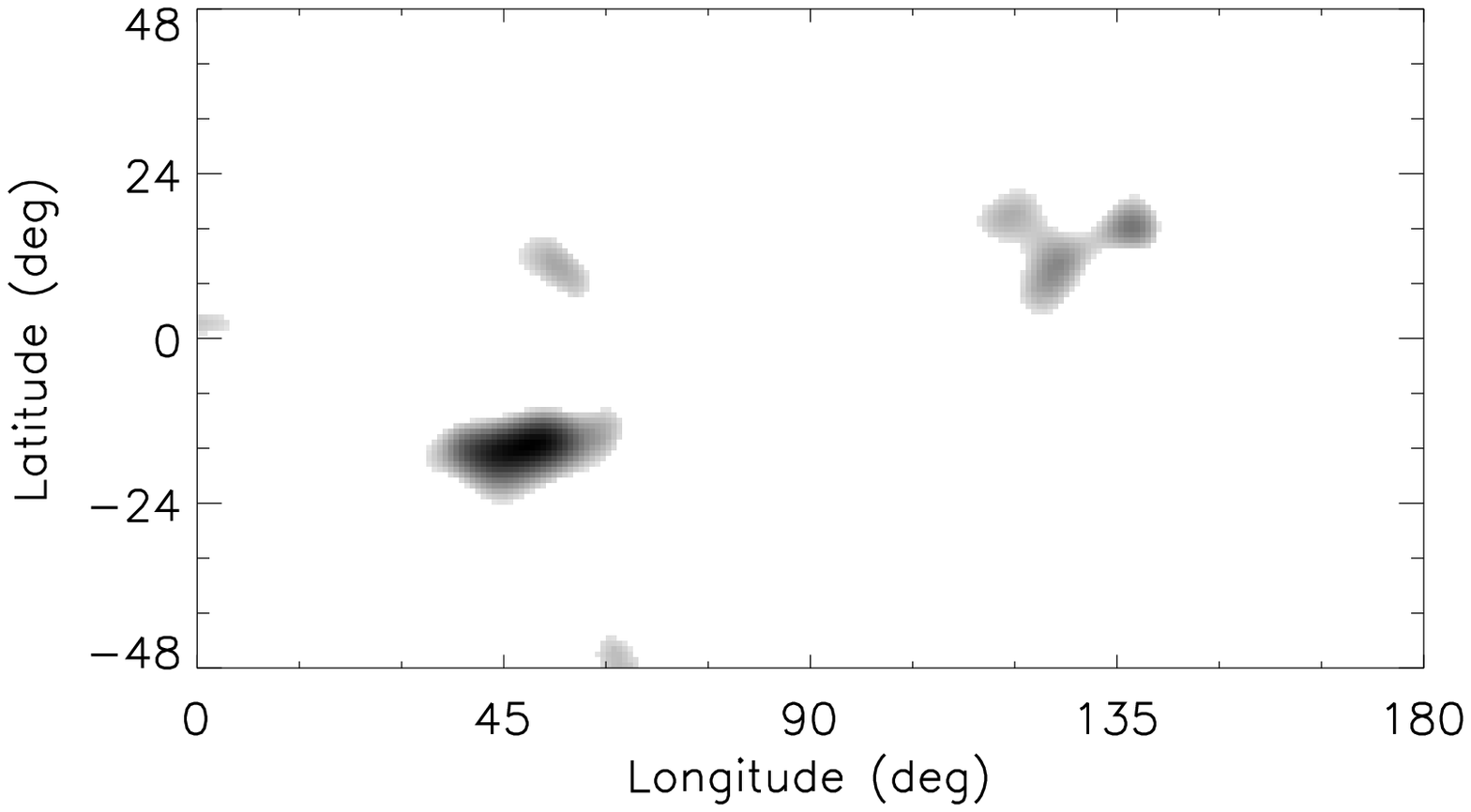}
              }
     \vspace{-0.35\textwidth}   % Shift close to the panel top
     \centerline{\Large \bf     % Includes the labels (here needs the color package)
      \hspace{0.0 \textwidth} \color{white}{(c)}
      \hspace{0.415\textwidth}  \color{white}{(d)}
         \hfill}
     \vspace{0.31\textwidth}    % Shift back to the panel bottom

\caption{Results of time-distance far-side active region imaging,
obtained from three-skip measurements (first row), combination of
four- and five-skip measurements (second row), combination of
three-, four-, and five-skip measurements by simple averaging (third
row) and combination of three-, four-, and five-skip measurements by
the new method of averaging (bottom row). On the right-side, images
display the far-side map after applying a travel-time threshold of
$-2.0 \; \sigma$ ($-1.5 \; \sigma$ for the three-skip map because
the standard deviation is substantially larger) in order to
highlight the far-side active regions that are of our interest.
Images are obtained for 8 November 2003.
        }
   \label{F-4panels}
   \end{figure}

\begin{figure}    %%%%%%%%%%%%%%%%%% FIGURE 2
                                % includes the two top panels
   \centerline{\hspace*{0.015\textwidth}
               \includegraphics[width=0.515\textwidth,clip=]{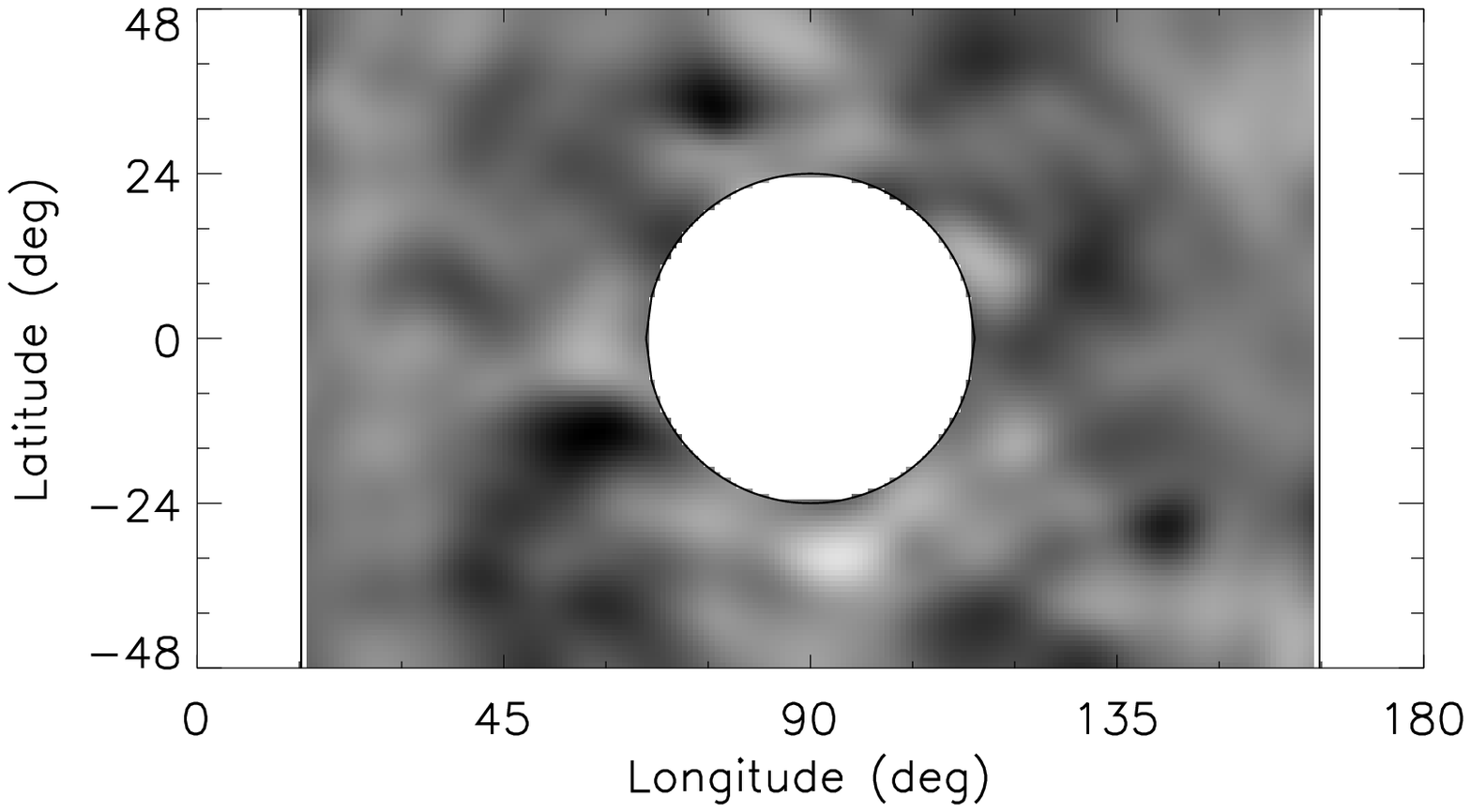}
               \hspace*{-0.03\textwidth}
               \includegraphics[width=0.515\textwidth,clip=]{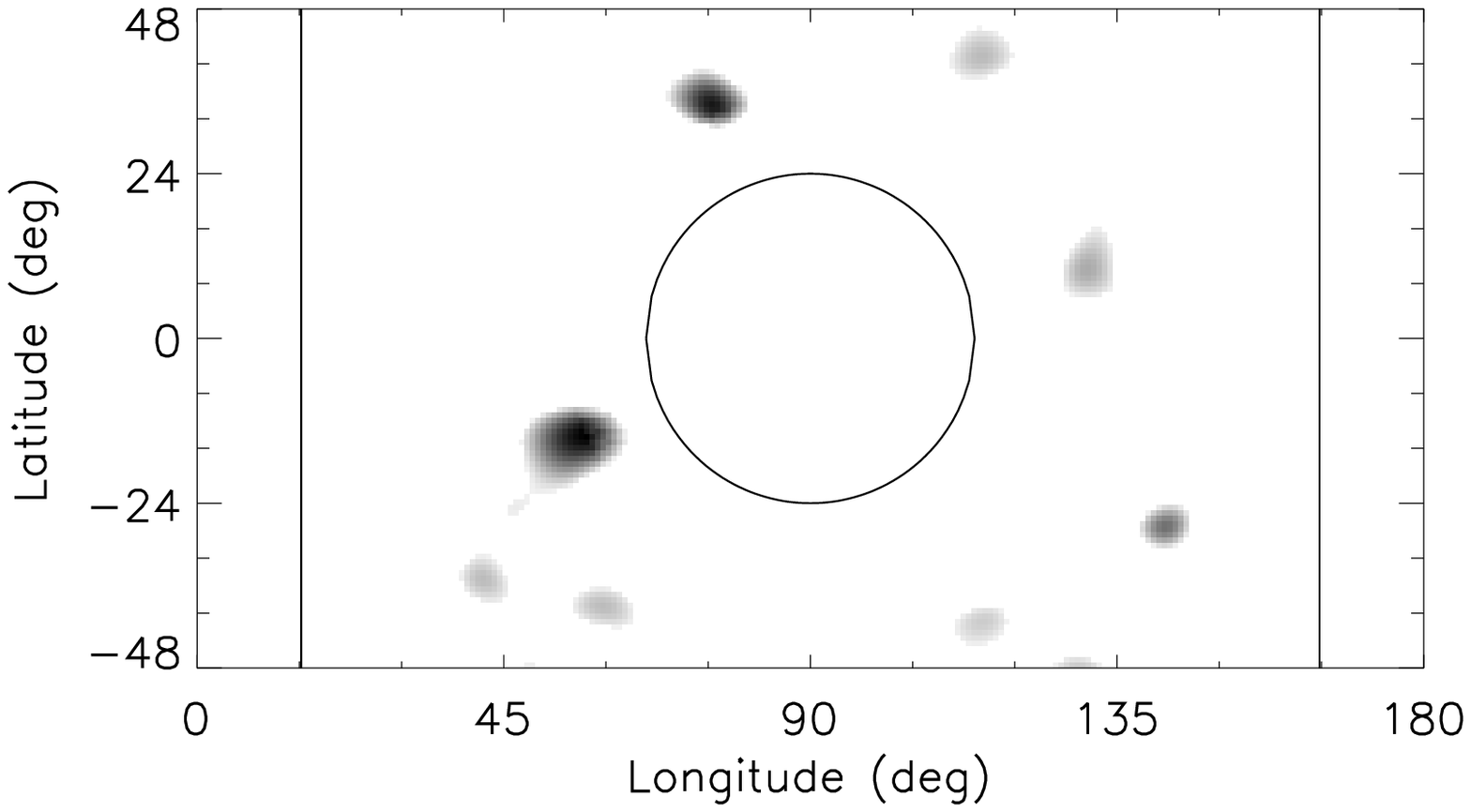}
              }
     \vspace{-0.35\textwidth}   % Shift close to the panel top
     \centerline{\Large \bf     % Includes the labels (here needs the color
                                %   package, see beginning of this file)
      \hspace{0.0 \textwidth}  \color{white}{(a)}
      \hspace{0.415\textwidth}  \color{white}{(b)}
         \hfill}
     \vspace{0.23\textwidth}    % Shift back to the panel bottom
   \centerline{\hspace*{0.015\textwidth}
               \includegraphics[width=0.515\textwidth,clip=]{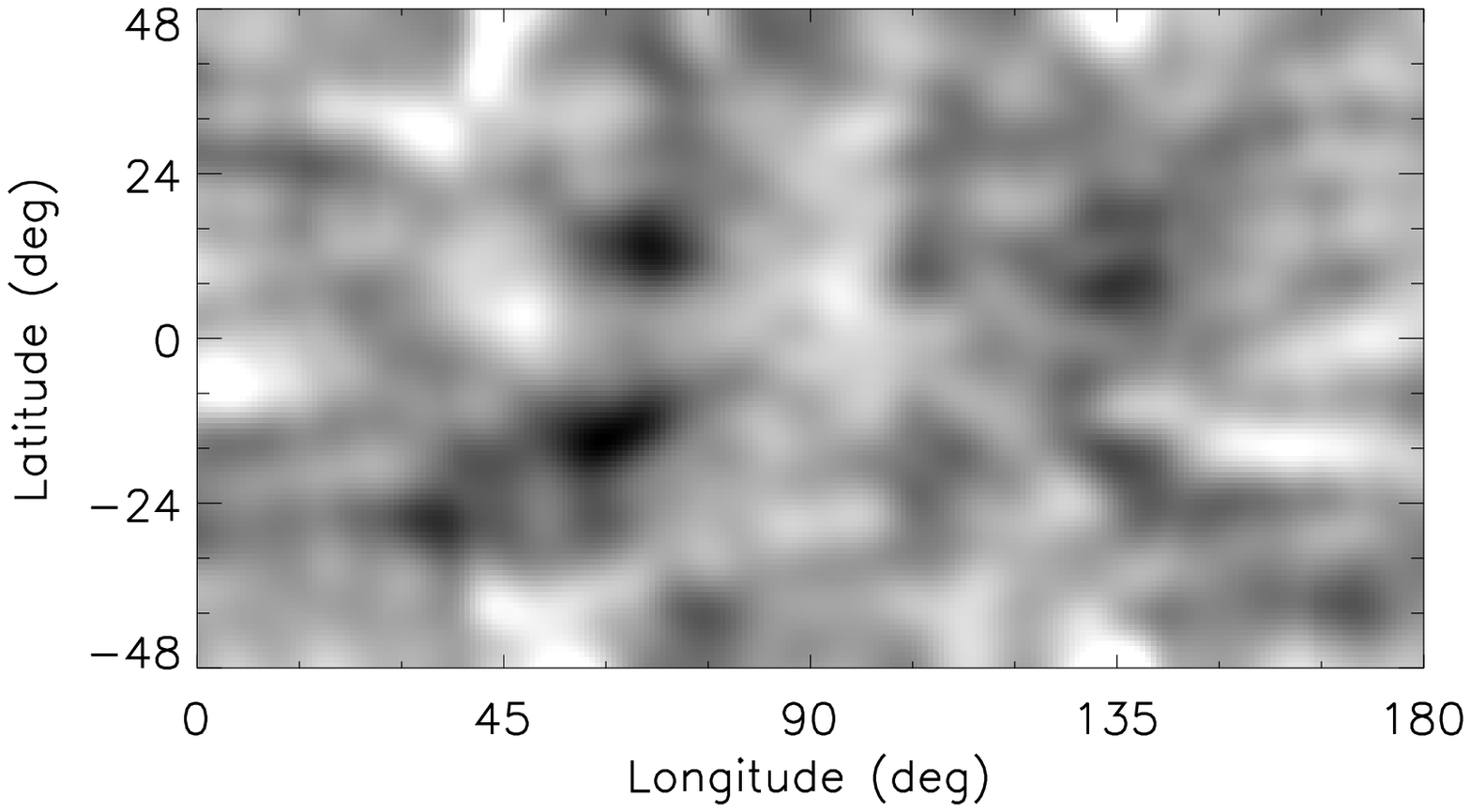}
               \hspace*{-0.03\textwidth}
               \includegraphics[width=0.515\textwidth,clip=]{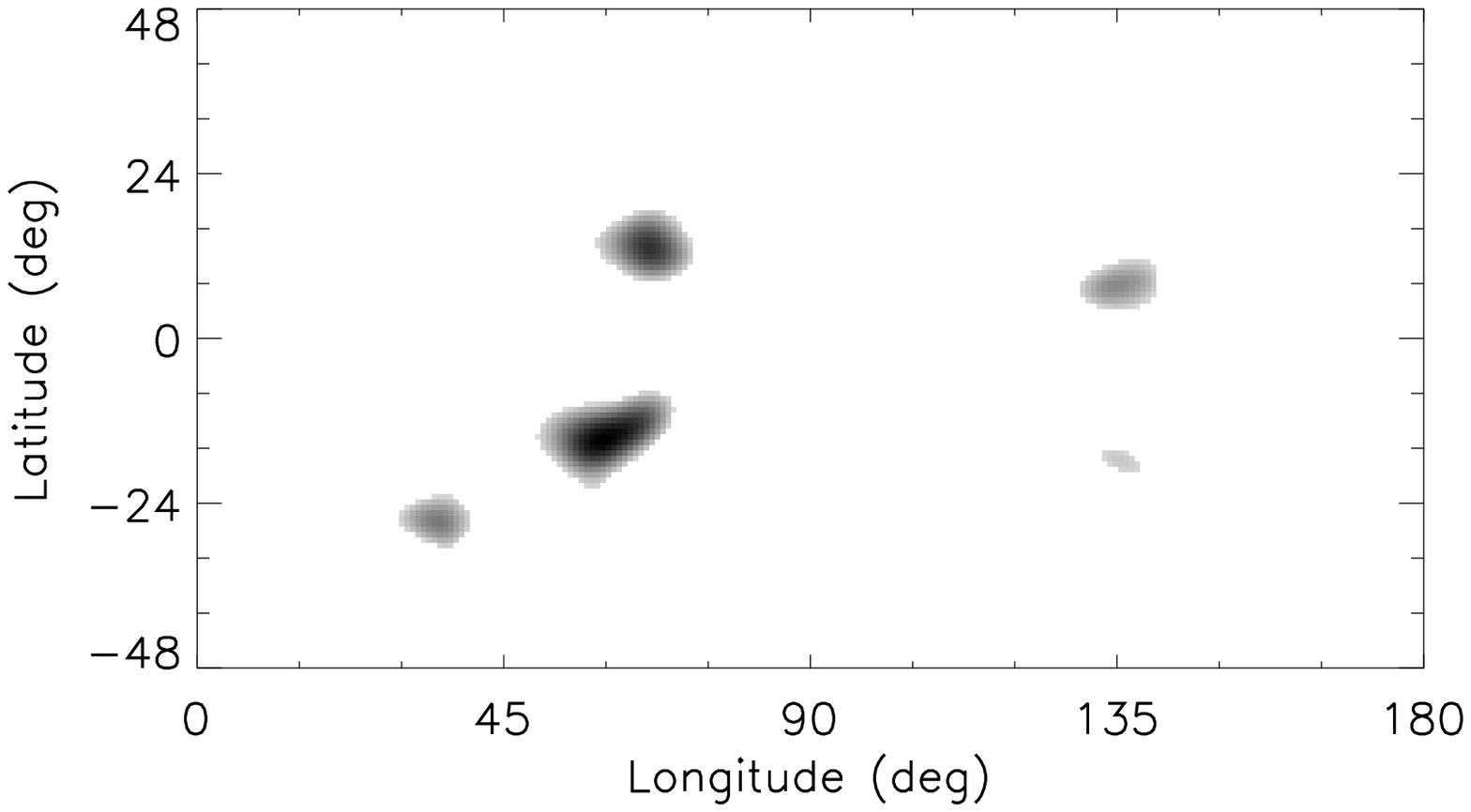}
              }
     \vspace{-0.35\textwidth}   % Shift close to the panel top
     \centerline{\Large \bf     % Includes the labels (here needs the color package)
      \hspace{0.0 \textwidth} \color{white}{(c)}
      \hspace{0.415\textwidth}  \color{white}{(d)}
         \hfill}
     \vspace{0.23\textwidth}    % Shift back to the panel bottom
   \centerline{\hspace*{0.015\textwidth}
               \includegraphics[width=0.515\textwidth,clip=]{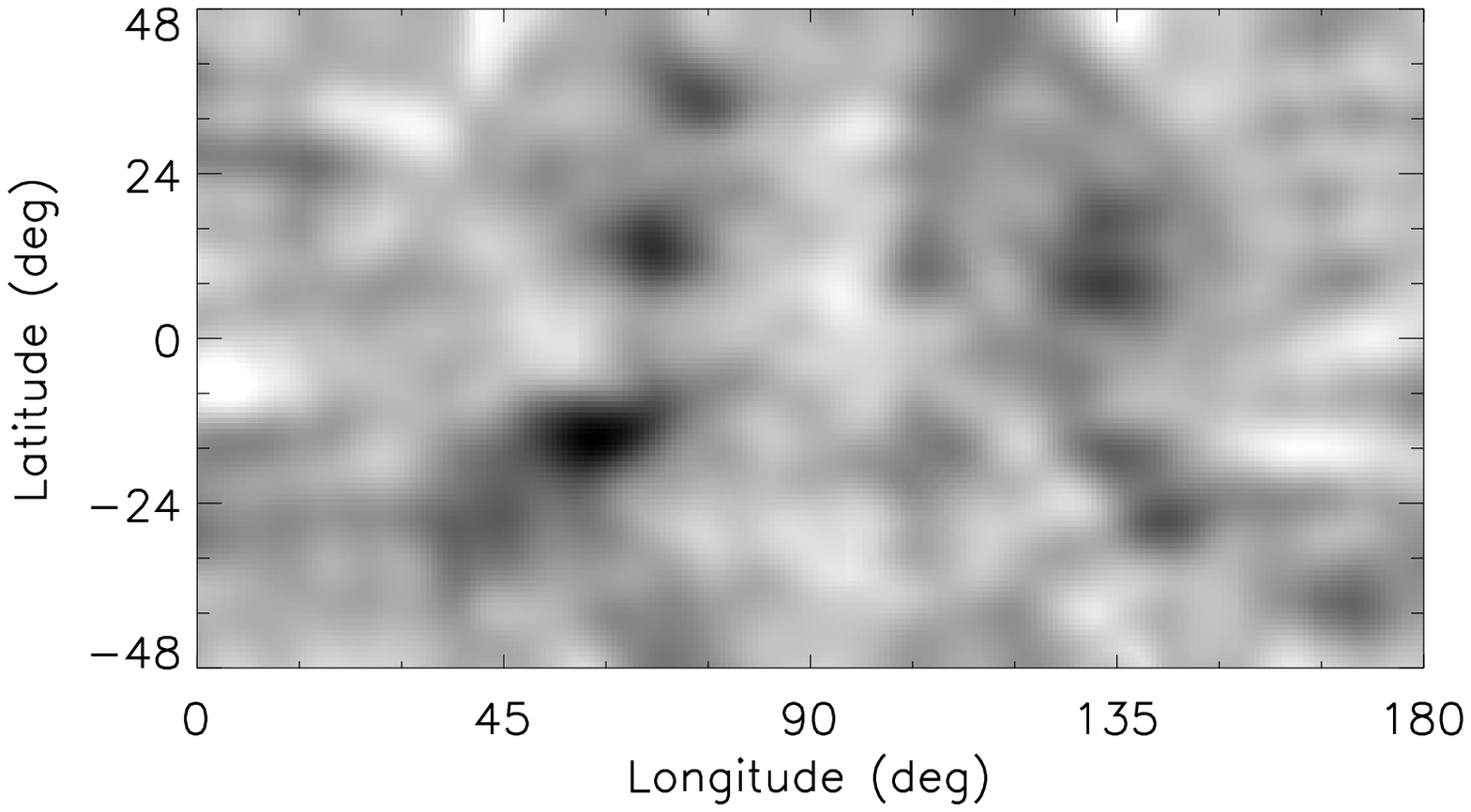}
               \hspace*{-0.03\textwidth}
               \includegraphics[width=0.515\textwidth,clip=]{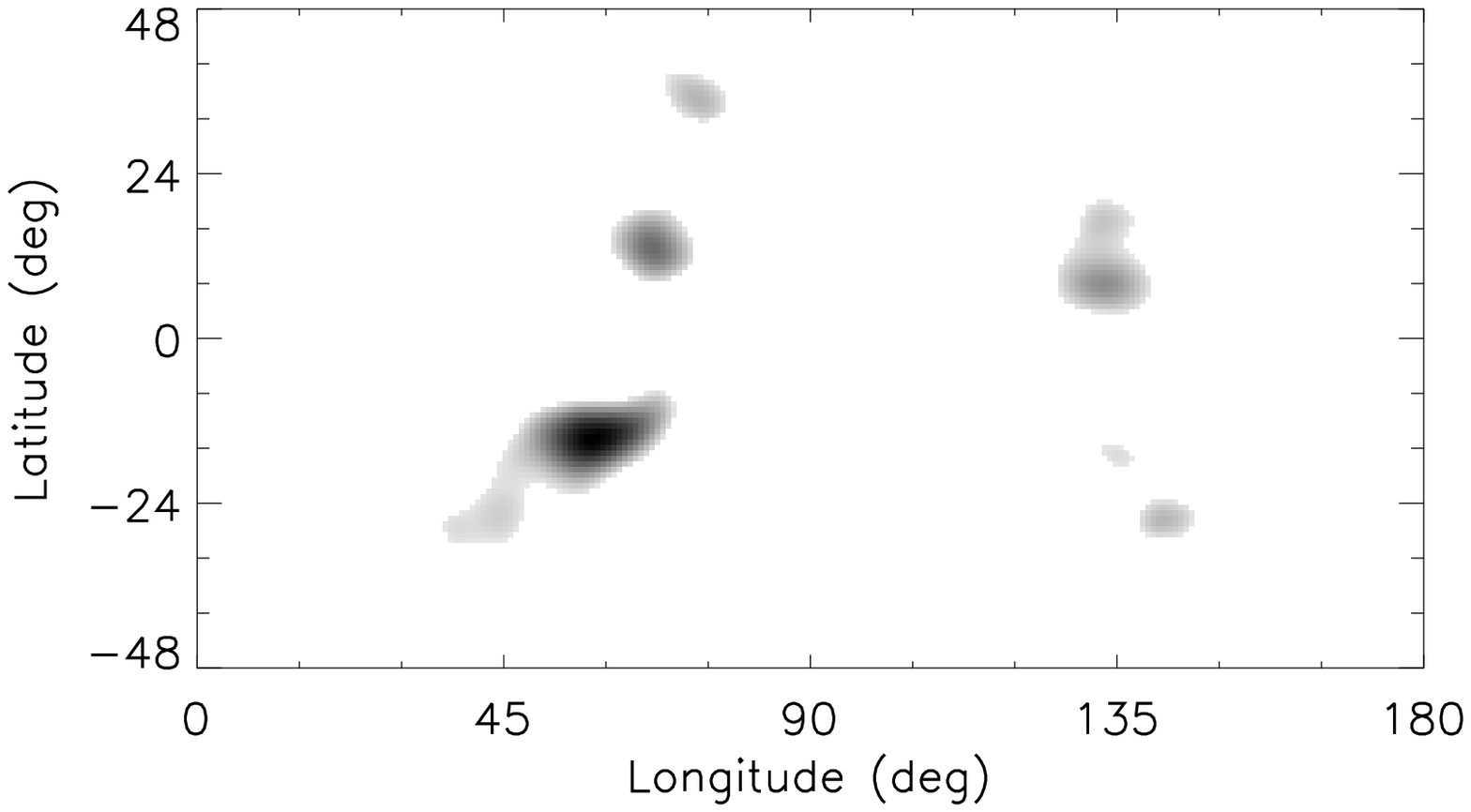}
              }
     \vspace{-0.35\textwidth}   % Shift close to the panel top
     \centerline{\Large \bf     % Includes the labels (here needs the color
                                %   package, see beginning of this file)
      \hspace{0.0 \textwidth}  \color{white}{(a)}
      \hspace{0.415\textwidth}  \color{white}{(b)}
         \hfill}
     \vspace{0.23\textwidth}    % Shift back to the panel bottom
   \centerline{\hspace*{0.015\textwidth}
               \includegraphics[width=0.515\textwidth,clip=]{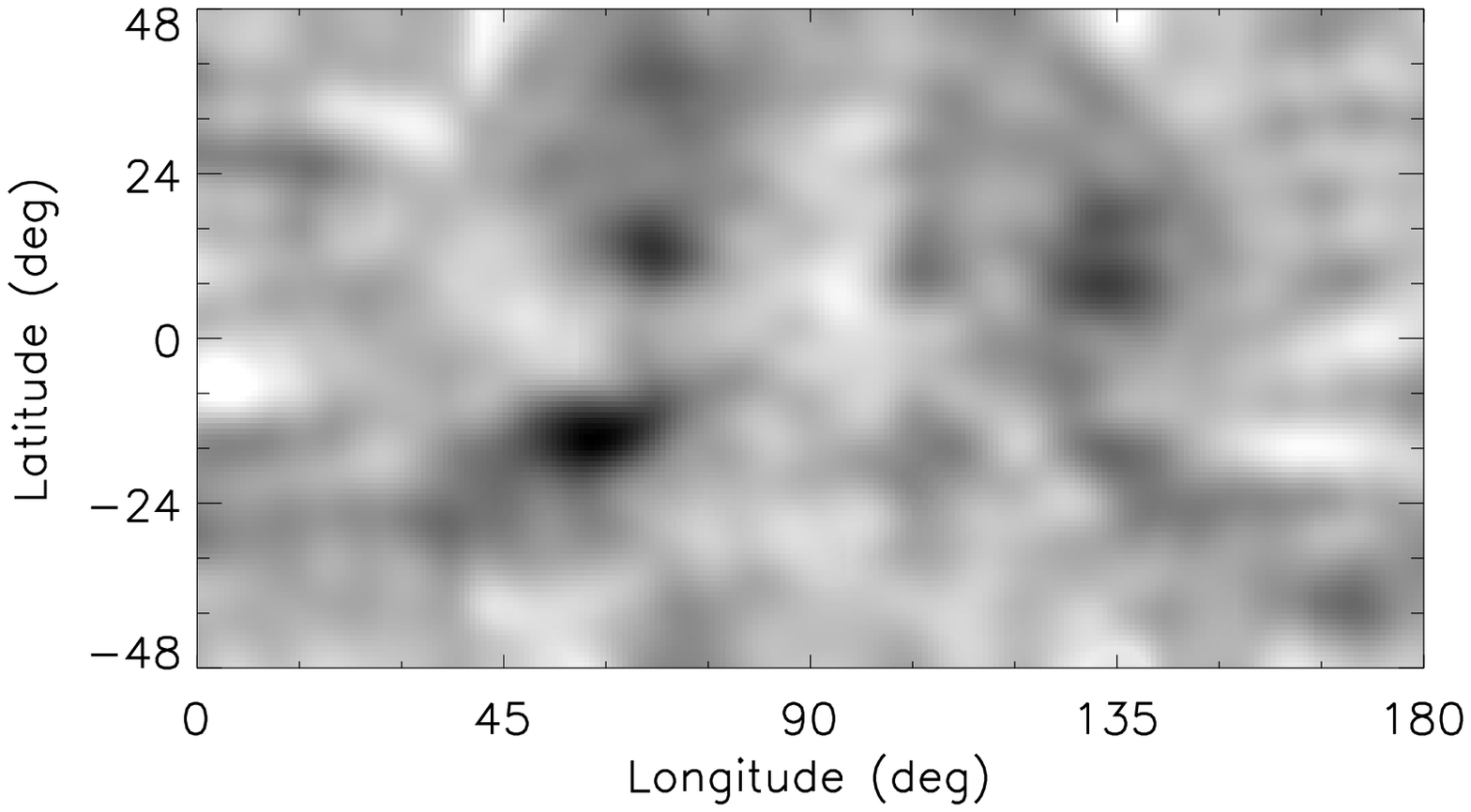}
               \hspace*{-0.03\textwidth}
               \includegraphics[width=0.515\textwidth,clip=]{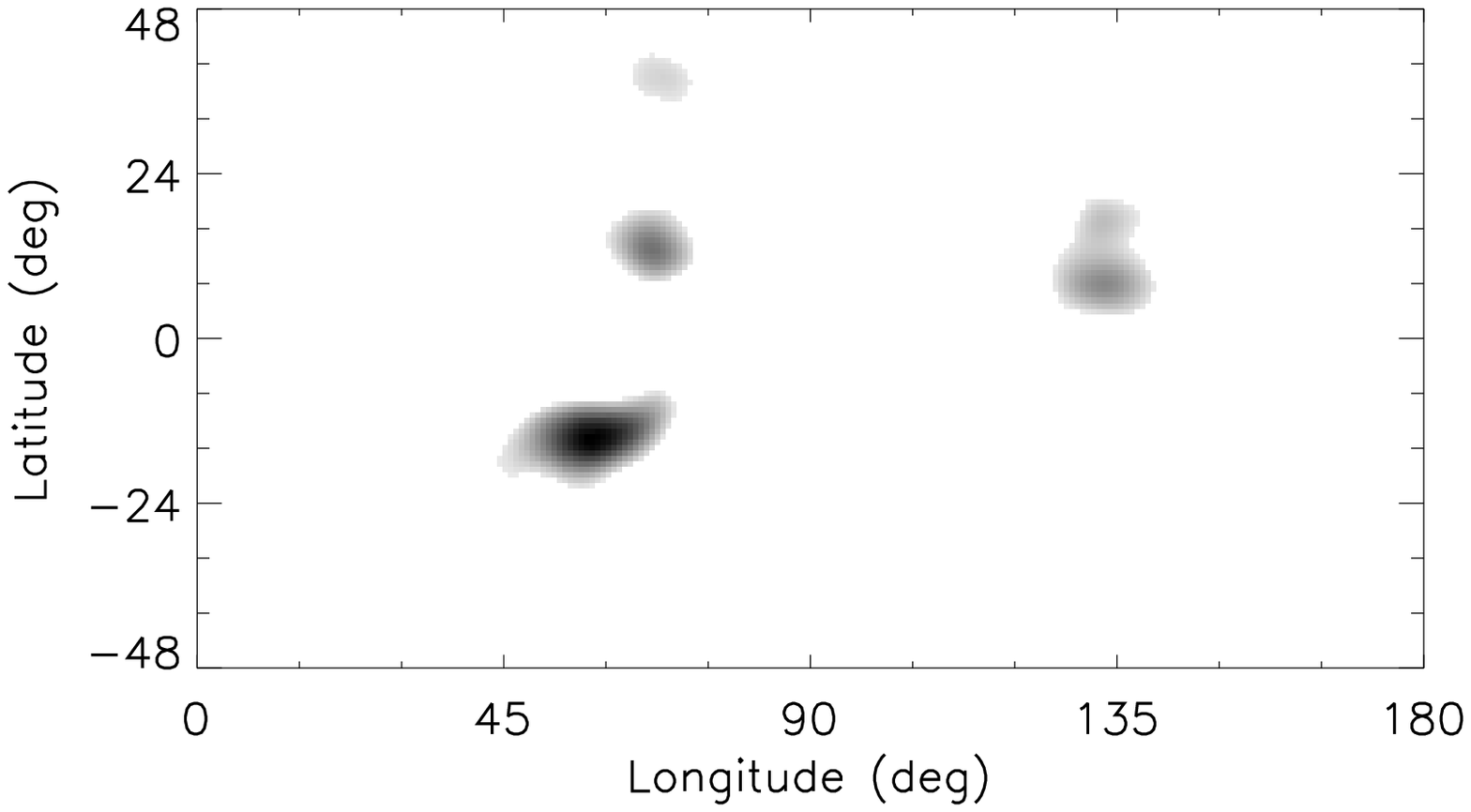}
              }
     \vspace{-0.35\textwidth}   % Shift close to the panel top
     \centerline{\Large \bf     % Includes the labels (here needs the color package)
      \hspace{0.0 \textwidth} \color{white}{(c)}
      \hspace{0.415\textwidth}  \color{white}{(d)}
         \hfill}
     \vspace{0.31\textwidth}    % Shift back to the panel bottom

\caption{Same as Figure 4, but results for 9 November, 2003.
        }
   \label{F-4panels}
   \end{figure}

\par Next we make far-side images using MDI
observational data. Two different datasets, each lasting 2048
minutes, with the middle time corresponding to 12:00 UT of 8 and 9
November 2003, are used for this purpose, and the corresponding maps
are shown on the left side of Figures 4 and 5 respectively. In order
to highlight the far-side active regions, the same maps are shown on
the right side of Figures 4 and 5 after applying a travel time
threshold of $-2.0 \; \sigma$. The threshold for the three-skip
image is smaller ($-1.5 \; \sigma$) because the standard deviation
of this travel time map is substantially larger than other maps
averaged from two or three travel time maps. In the first row of
Figures 4 and 5, far-side images made with the new three-skip
technique are displayed. In order to test the quality and accuracy
of the new imaging technique, far-side images from the same dates
made with the existing time-distance technique (Zhao, 2007) are
shown in the second row. The third and fourth rows display far-side
images made with a combination of three-, four-, and five-skip
results, but the technique applied for the combination is different
in each case, and it is further discussed in more details below.
\par The new far-side
maps that utilize acoustic signals with three, four, and five skips
consist of three distinct regions, two bands along the solar limbs
$15^\circ$ in width, a circular area around the center of the far
side with a radius of $24^\circ$, and the rest of the far-side
region. The imaging of the bands along the solar limbs is based only
on acoustic signals with four skips. A combination of four- and
five-skip acoustic signals is used for the central circular area,
while for the rest of the far-side image the combination includes
three-skip measurements in addition to the four- and five-skip
measurements. The imaging of the existing far-side maps is based on
a combination of four- and five-skip measurements except for two
bands along the solar limbs where only four-skip measurements are
used (see Zhao, 2007). The boundaries of all of those regions have
been linearly smoothed over a distance of $12^\circ$. All of the
far-side acoustic travel times are displayed after a mean travel
time background is removed.

\par As mentioned before, the two bottom rows of Figures 4 and 5 present
far-side images made by combining three-, four- and five-skip
measurement schemes. The method of combining images by simply taking
the average is a rather simple but reasonable technique in the case
of two images and is able to enhance the signal-to-noise ratio in
the far-side images (Zhao, 2007). Therefore, the arithmetic average
is used in the central circular area of the far side that involves
only acoustic signals with four and five skips. However, more
sophisticated techniques of combination can be applied if three or
more far-side images are available. At each pixel location of the
far-side map where there is overlap between the regions accessed by
the different combinations of skip distances, the travel times
obtained from the three-, four-, and five-skip images are considered
and the absolute values of the three differences in the travel times
are calculated. If the two largest differences are larger than a
high threshold and the smallest difference is smaller than a low
threshold then the travel time that is substantially smaller or
larger is replaced by the average of the other two. The travel time
at those pixels that satisfy the above conditions is essentially the
average of travel times of just two different measurement schemes.
By setting the high threshold equal to 13 seconds and the low
threshold equal to 9 seconds we have managed to remove spurious
features from the far-side map and at the same time enhance the
signal in the active regions. The third and fourth rows of Figures 4
and 5 show respectively far-side maps made by simply averaging the
three-, four-, and five-skip acoustic signals and by applying the
new combination technique.

\par Both the new and the existing far-side
imaging technique cannot give consistently a clear far-side map
without any spurious features but they can differ quantitatively as
well as qualitatively. It is clear from Figure 4 and 5 that the
level of noise in the three-skip image is quite high. If we simply
average the three-, four-, and five-skip signals and compare the new
far-side map with the existing map, the large active regions appear
larger and more enhanced but spurious signals originating from the
contribution of the noisy three-skip image also appear in the
far-side map. However, the new combination technique is able either
to completely remove some of those signals from the far-side map or
at least to weaken their contribution. As a result the new far-side
maps have a better signal-to-noise ratio with reduced and fewer
artifacts.

\par Comparing corresponding maps we can conclude that the detection
of large active regions is more successful with the new technique.
The three-skip measurement scheme utilizes low-$\ell$ acoustic
modes, lower than the ones used in the four- and five-skip schemes
hence the wavelength of these modes is greater, and this results in
a reduced sensitivity in the detection of small active regions. But
even if the three-skip scheme completely failed to detect small
active regions, the new combination technique ensures that in most
cases these regions will be visible in the far-side map. On the
other hand, increasing the total number of skips produces
degradation in the signal correlations and, as we expect, the
three-skip method has stronger correlations than the four- and the
five-skip method. It is quite remarkable though that in some cases
the three measurement schemes have spurious signals at the same
locations, an indication that there might be some weak correlation
in the noise of the far-side images, or alternatively, that
``ghost-images" from near-side active regions appear on the far
side.

\section{Discussion}

\par The availability of both time-distance and helioseismic
holography far-side maps makes the detection of active regions more
robust and confident. Additionally, far-side imaging provides us
with some experience in analyzing low- and medium-$\ell$ mode
oscillations by the use of local-helioseismology techniques and this
will help us in analyzing deeper solar interior and polar areas.

\par The time-distance technique with three-skip acoustic signals is
another imaging tool in addition to the existing techniques. We have
successfully made  three-skip far-side images of the solar active
regions using data from numerical simulations as well as from MDI
observations. The three-skip measurement scheme cannot cover the
whole far side, and moreover, the far-side images have many spurious
features that do not appear in the four- and five-skip images.
Consequently the three-skip images by themselves are not as good as
the images obtained by other time-distance techniques. Our
motivation for making three-skip far-side images derives from the
fact that the availability of one more far-side imaging technique
makes possible the application of more sophisticated methods of
combining images than the simple arithmetic averaging. In addition,
the correlations in the three-skip technique are rather strong and
facilitate the detection of large active regions.

\par The combination of four- and five-skip measurement schemes
was shown by Zhao (2007) to significantly enhance the
signal-to-noise ratio of far-side acoustic travel time measurements
and make the resulting far-side map much cleaner. More specifically
it helps to remove most but not all of the spurious features. On the
other hand it also removes some small active regions that can
otherwise be seen in one map or the other. The three-skip scheme is
an independent tool for imaging the far-side solar active regions
and thus it can be combined with the two existing time-distance
techniques. The combination of the three measurements can give a
cleaner far-side map, with fewer spurious features and enhanced
active regions.

\par The method of combining images is very important in the problem
of far-side imaging. In this paper we present two different methods,
the first one was also used by Zhao (2007) while the second one is a
newly suggested method that improves the quality of the image. The
new method utilizes, at each pixel location, the far-side maps
derived from three-, four-, and five-skip travel times, subject to
two arbitrary thresholds. Properly adjusting these thresholds it is
possible to enhance the active regions and reduce the level of
noise.

%% Figure
%
% \begin{figure}
% \centerline{\includegraphics[width=0.5\textwidth,clip=]{<fig.eps>}}
% \caption{}%\label{fig:?}
% \end{figure}

%% Table
%
% \begin{table}
% \caption{}%\label{tbl:?}
% \begin{tabular}{}
% \hline
% \multicolumn{2}{c}{<>}
% <data>
% \hline
% \end{tabular}
% \end{table}

%%%%%%%%%%%%%%%%%%%%%%%%%%%%%%%%%%%%%%%%%%%%%%%%%%%%%%%%%%%%%%%%%%%%%%%%%%%
%% Appendix
%
% \appendix

%%%%%%%%%%%%%%%%%%%%%%%%%%%%%%%%%%%%%%%%%%%%%%%%%%%%%%%%%%%%%%%%%%%%%%%%%%%
%% Acknowledgements
%
\begin{acks}
The numerical simulations used in $\!$ this paper were performed at
NASA Ames Research Center. T.H. was supported by the NASA Living
With a Star program and the NASA Postdoctoral Program administered
by Oak Ridge Associated Universities under contract with NASA.
\end{acks}

\end{article}
\end{document}